\documentclass[11pt]{article}
\usepackage{moriond,epsfig}

\def\Journal#1#2#3#4{{#1} {\bf #2}, #3 (#4)}

\def\NPB{{\em Nucl. Phys.} B}
\def\PLB{{\em Phys. Lett.}  B}
\def\PRL{\em Phys. Rev. Lett.}
\def\PRD{{\em Phys. Rev.} D}

\def\POS{{\em PoS}}

\def\be{\begin{equation}}
\def\ee{\end{equation}}
\def\bea{\begin{eqnarray}}
\def\eea{\end{eqnarray}}

\begin{document}
\vspace*{4cm}
\title{QCD THERMODYNAMICS WITH ALMOST REALISTIC QUARK MASSES}

\author{ C.~Schmidt\\(for the RBC-Bielefeld Collaboration) }

\address{Department of Physics, Brookhaven National Laboratory, Upton 11973 NY, USA}

\maketitle\abstracts{
Ongoing calculations on the QCDOC supercomputer at Brookhaven National
Laboratory and the APEnext installation at the University of
Bielefeld aim to determine the critical temperature of the QCD phase
transition as well as the equation of state with almost realistic
quark masses. We will discuss preliminary results of the quark mass
and cut-off dependence of order parameters, susceptibilities, static
quark potentials and the critical temperature in (2+1)-flavor QCD. All
these quantities are of immediate interest for heavy ion phenomenology.
}

\section{Introduction and Lattice Setup}
The calculations of the QCD phase diagram and bulk 
thermodynamic quantities from first principle give input for Heavy Ion 
Phenomenology, Cosmology and Astrophysics. In particular, it is mandatory to
improve estimates on the critical temperature ($T_c$) to make contact with HIC
Phenomenology. Since for the critical energy density ($\varepsilon_c$) we have 
$\varepsilon_c \sim T_c^4$, a small error on $T_c$ is important. Moreover, an
interesting question is whether or not the freeze-out temperature in HICs is
connected to $T_c$.

The grand canonical QCD partition function on the lattice is given by the
integral
\begin{equation}
Z_{GC}(am_q, am_s, N_s, N_t, \beta)=\int {\cal D}U\; 
\left[\det M(U,m_q)\right]^2\left[\det M(U,m_s)\right]\exp\{-\beta
S_G[U]\}\quad ,
\end{equation}
where $a$ is the lattice spacing, $am_q$ and $am_s$ are the light and strange
quark masses respectively (given in lattice units), $N_s$ and $N_t$ are the
number of lattice points in spacial and temporal direction. $M$ is the fermion
matrix and $U$ are the gauge fields, located on the links between the lattice
points. $S_G$ is the gauge part of the action and $\beta$ is the coupling which
controls the lattice spacing. The lattice action we use is especially designed
for finite temperature QCD Simulations, where the lattice spacing is
usually rather large. In the gauge sector we use a $(2\times1)$-Symanzik
improvement scheme which eliminates all cut-off effects of order ${\cal
  O}(a^2)$. For the fermions we use the staggered fermion formulation. On top
of that we add an improvement term which 
restores the rotational symmetry of the free quark propagator on the lattice up
to order ${\cal O}(p^4)$ in the momentum \cite{Heller:1999xz}. In order to
improve the flavor symmetry, which is violated in the staggered fermion
formulation, we smear each link of the gauge field which is used in the
standard part of the fermion action by its surrounding three link staples
(p4fat3). The p4fat3 action was used for thermodynamical calculation earlier
\cite{Karsch:2000ps,Karsch:2000kv}, reports of the ongoing project
have been given by C.~Jung \cite{p4fat7_1} and M.~Cheng \cite{p4fat7_2}

We perform simulations with 2 light and one heavy quark flavor. The strange
quark mass is always fixed to the physical value, whereas the light quark mass
is varied in the range of $m_q=0.05m_s-0.4m_s$. The lattices have temporal extent
$N_t=4,6$, which corresponds at the critical temperature ($T_c$) to a lattice
spacing of $a\approx0.13$~fm and $0.22$~fm respectively. The lattice extent in
spacial direction is $N_s=8,16,32$. To determine the scale, we perform zero 
temperature simulations on $16^3\times 32$ lattices. These calculations are
being performed on the QCDOC  computers at BNL and the APEnext
installation at the University of Bielefeld.

\section{Order Parameters and Susceptibilities}
\begin{figure}
\begin{minipage}{.48\textwidth}
\begin{center}
\includegraphics[width=.99\textwidth]{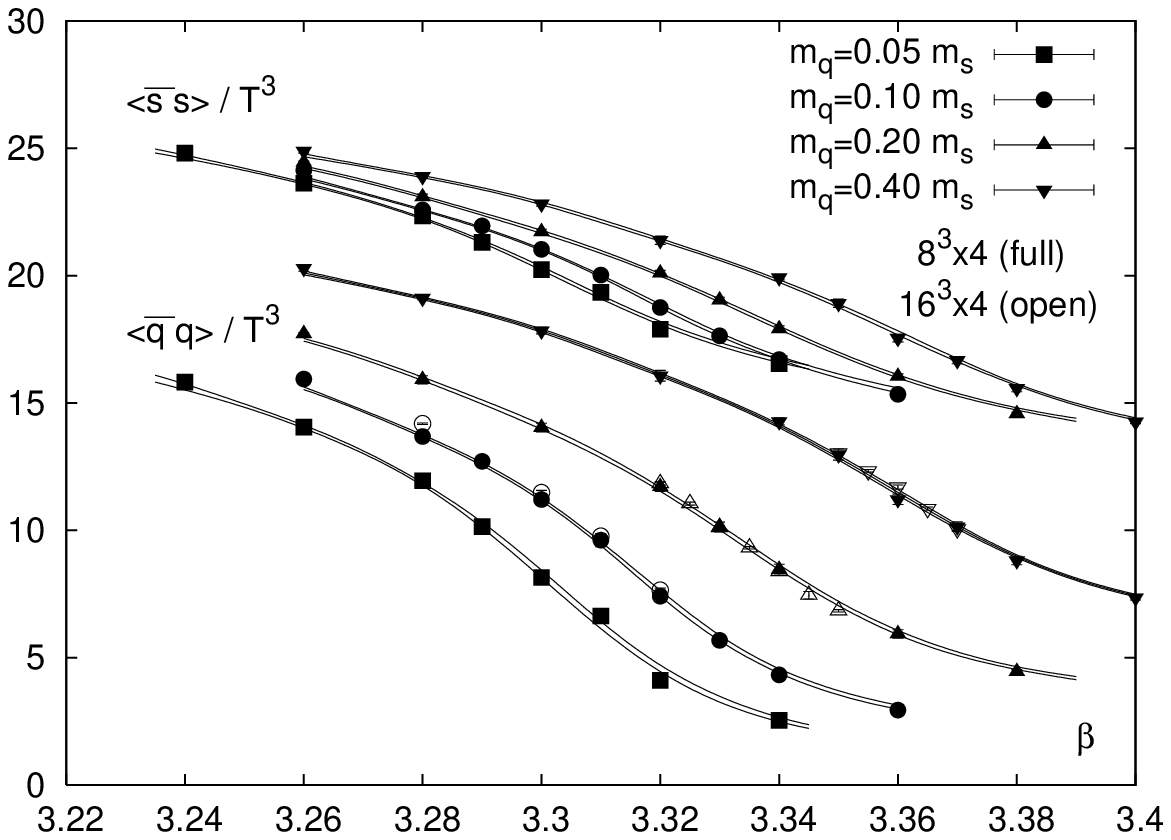}
\end{center}
\end{minipage}
\begin{minipage}{.48\textwidth}
\begin{center}
\includegraphics[width=.99\textwidth]{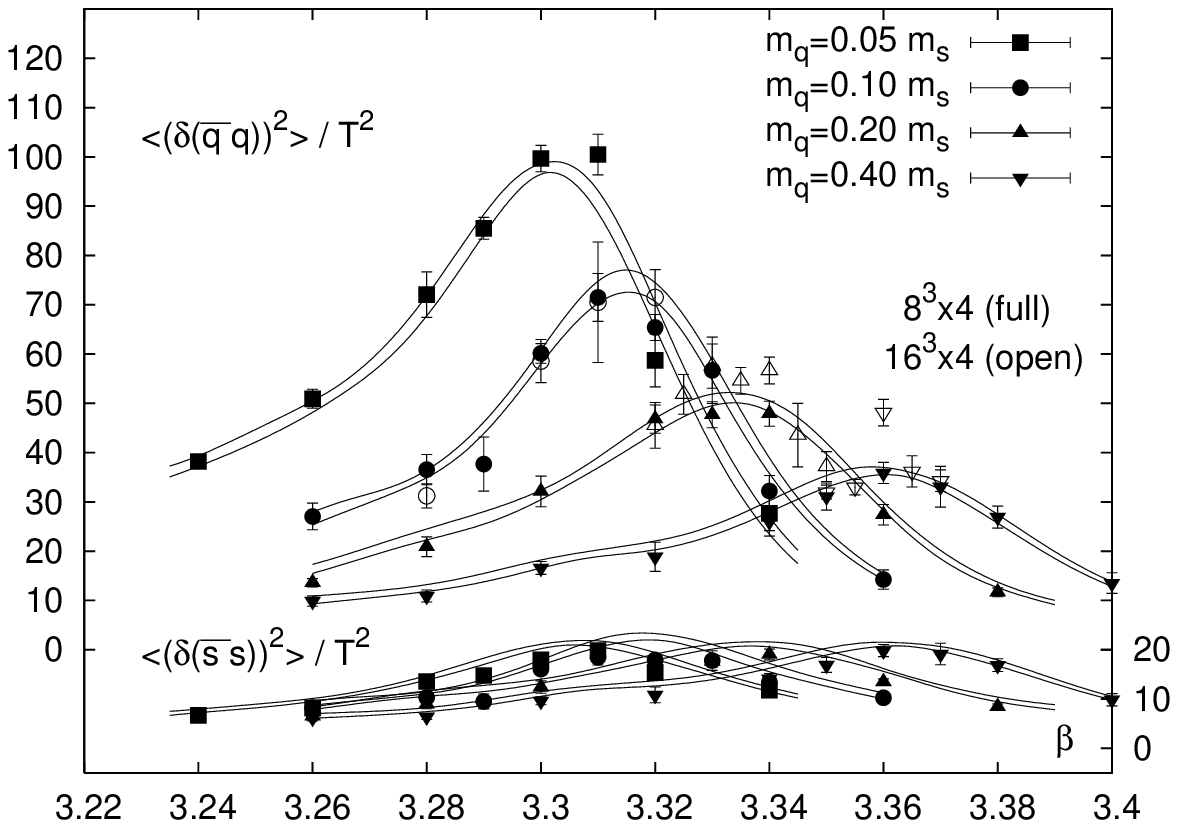}
\end{center}
\end{minipage}
\caption{The light quark and strange quark chiral condensate at (left)
  and its susceptibility (right) as a function of the
  coupling. The strange quark chiral condensate was multiplied with a factor 2
  for better visibility. Calculation have performed with various values of the
  light quark mass but fixed strange quark mass. Lattice sizes are
  $8^3\times 4$ and $16^3\times 4$.}\label{fig:chiral}
\end{figure}

Connected to chiral symmetry breaking is the chiral condensate. In the staggered
formulation of lattice QCD it is given by
\begin{eqnarray}
\left<\bar q q\right> &\equiv& \left.
\frac{{\rm}d}{{\rm d}m}\ln Z_{GC}(am, am_s, N_s, N_t, \beta)
\right|_{m=m_q}=N_s^{-3}N_t^{-1}\frac{1}{2}\left<{\rm Tr}M_{KS}^{-1}(m_q)\right> \quad ,\\
\left<\bar s s\right> &\equiv& \left.
\frac{{\rm}d}{{\rm d}m}\ln Z_{GC}(am_q, am, N_s, N_t, \beta)
\right|_{m=m_s}=N_s^{-3}N_t^{-1}\frac{1}{4}\left<{\rm Tr} M_{KS}^{-1}(m_s)\right> \quad ,
\end{eqnarray}
where $M_{KS}$ is the staggered fermion matrix. In the limit of vanishing quark
masses the chiral condensate is an exact order parameter of the spontaneous
chiral symmetry breaking. Its expectation value is non-zero below the critical
temperature ($T_c$) and zero above. At finite quark masses where the chiral
symmetry is explicitly broken, the chiral condensate still signals the
transition by a rapid change. In Fig.~\ref{fig:chiral} we show the expectation
value of the light quark chiral condensate ($\left<\bar q q \right>$) and
strange quark chiral condensate ($\left<\bar s s \right>$) as function of the
coupling $\beta$ (left) and its susceptibility 
(right) on $N_t=4$ lattices. 
Results on the chiral susceptibility on $N_t=6$ lattices are shown in
Fig.~\ref{fig:pot}(left). The results have been interpolated in the
coupling $\beta$ by using the multi-histogram re-weighting
technique \cite{Ferrenberg:1988yz}. Since the coupling controls the lattice
spacing $a$ it thus also controls the temperature, which is given by
$T=1/N_ta$. Large couplings $\beta$ corresponds to large temperatures $T$ and
small couplings to small temperatures. At the 
coupling where the condensate shows the most drastic change, the
corresponding susceptibility peaks. We define the critical coupling
$\beta_c$ by the peak position of the chiral susceptibility. In
Fig.~\ref{fig:chiral}, we
compare results for different quark masses. The calculation with
$m_q=0.05m_s$ is roughly at the physical point. A clear
dependence of the critical coupling (critical temperature) on the quark
mass can be observed. The peak positions of
the light quark and strange quark susceptibilities do however coincide
within our statistical accuracy. Which indicates that chiral symmetry
restoration for light and strange quark occurs at the same
temperature. The strength of the transition
decreases with increasing quark masses, this is reflected in a
decreasing peak height of the susceptibilities. 

In Fig.~\ref{fig:chiral} we also compare results form $8^3\times 4$ and
$16^3\times 4$ lattices. Since we see almost no volume dependence the
results suggest that the transition is in fact not a true phase
transition in the thermodynamic sense but a rapid crossover.

\section{Scale Setting and the Static Quark Potential}
\begin{figure}
\begin{minipage}{.48\textwidth}
\begin{center}
\includegraphics[width=.99\textwidth]{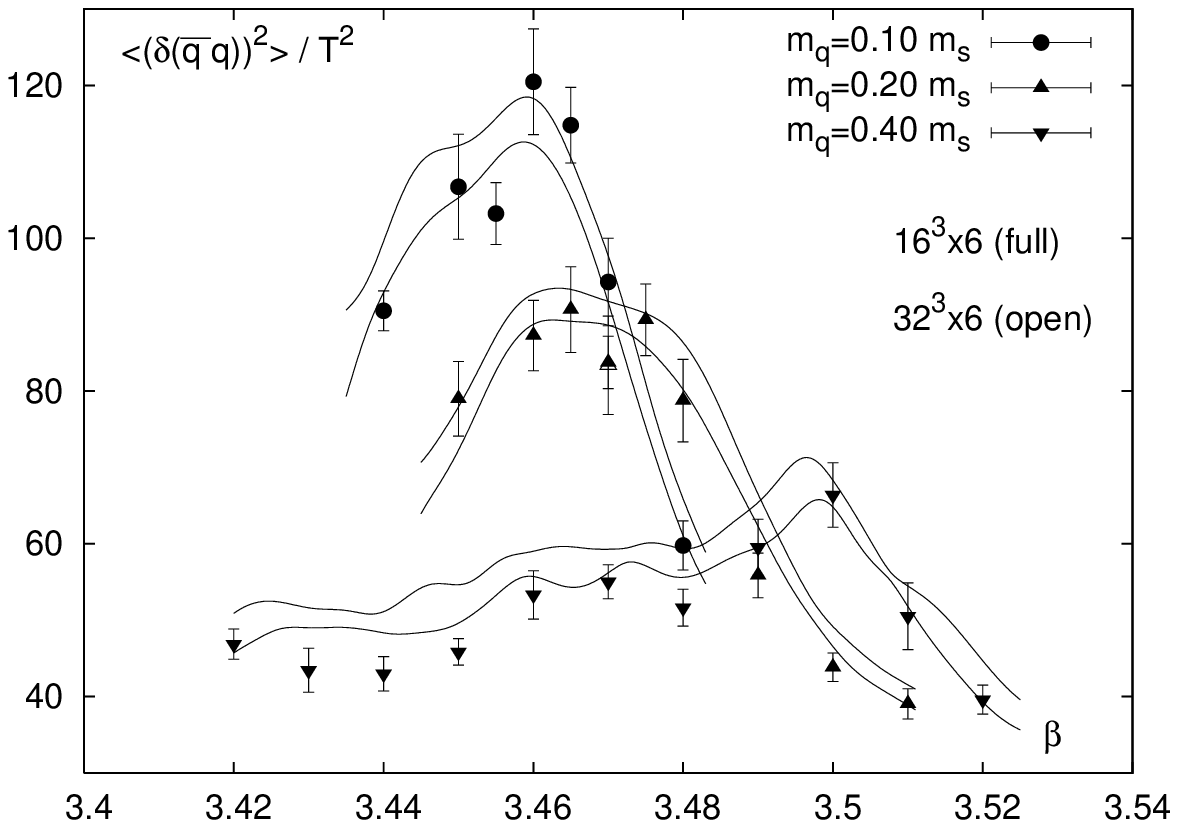}
\end{center}
\end{minipage}
\begin{minipage}{.48\textwidth}
\begin{center}
\includegraphics[height=.67\textwidth, width=.99\textwidth]{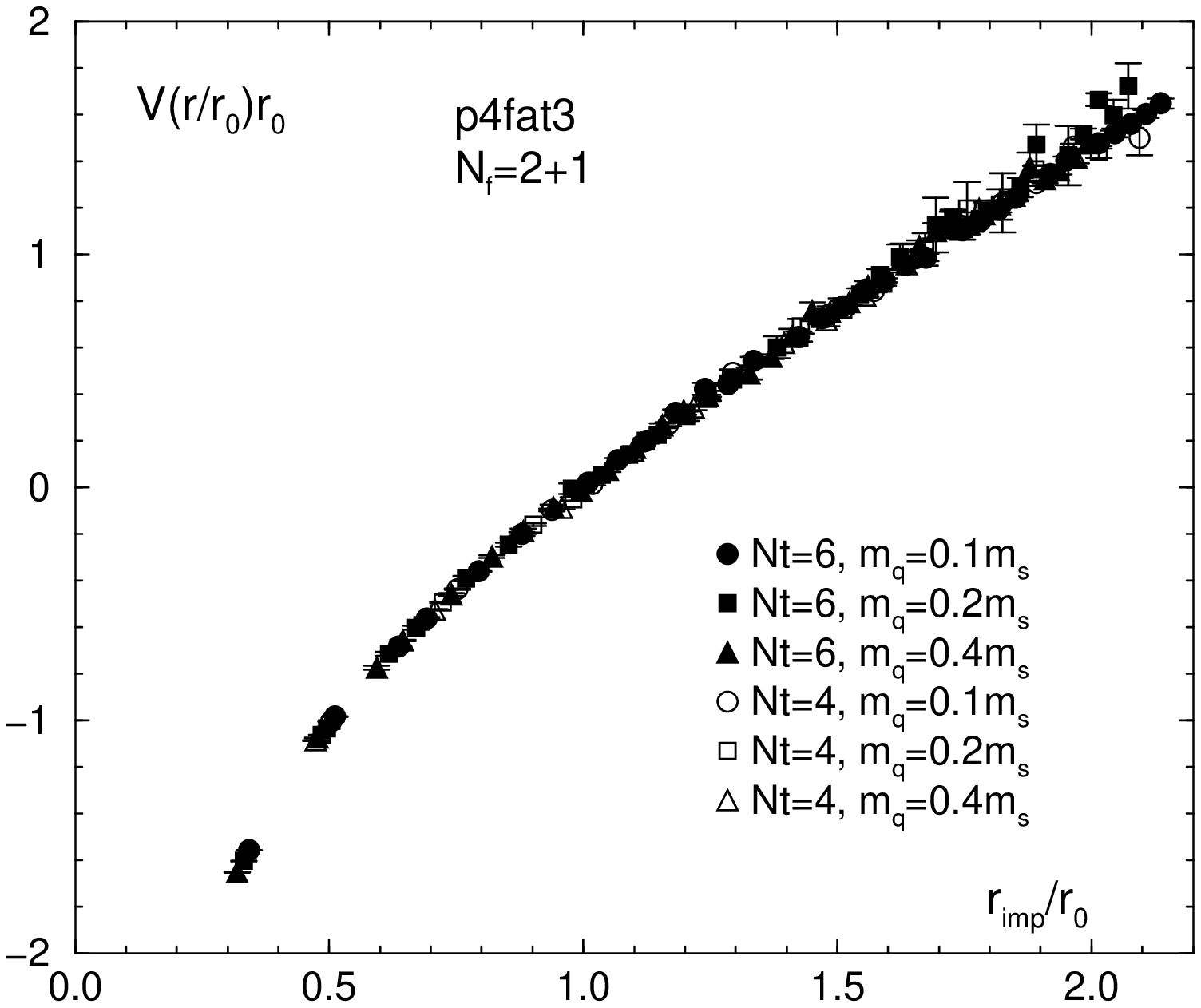}
\end{center}
\end{minipage}
\caption{The light quark chiral susceptibility as a function of the
  coupling measured on $16^3\times 6$ lattices with three
  different light quark masses (left). The zero temperature static
  quark potential in units of the Sommer scale. Results for different
  light quark masses and lattice spacings are plotted seem to fall on one
  universal curve.}\label{fig:pot}
\end{figure}

On $16^3\times 32$ lattices we calculate the zero temperature static
quark potential for all quark mass values at their corresponding
critical couplings. The Sommer scale $r_0$ is defined as the distance
where the derivative of the potential take a certain value, to be
precise it is defined as
\begin{equation}
\left. r^2\frac{{\rm d}V}{{\rm d}r}\right|_{r=r_0}=1.65\quad .
\end{equation} 
The Sommer scale is often used to set the scale in lattice
calculations, a phenomenological value is $r_0=0.5$~fm. In
Fig.~\ref{fig:pot} (right), we plot the static quark potential for
different quark masses and lattice spacing in units of the Sommer 
since all results seem to fall on a universal curve, the static quark
potential shows almost no quark mass or cutoff dependence. 
In order to remove short range lattice artifacts, we have used improved
distances $r_{imp}$ in the Coulombic part of the potential. For our choice of
the lattice action $r_{imp}$ is given by
\begin{equation}
\frac{1}{4\pi r_{imp}}\equiv \int\frac{d^3k}{(2\pi)^3}
\frac{e^{ikr}}{4\sum_i(\sin^2{\frac{k_i}{2}})
+\frac{1}{3}\sin^4{\frac{k_i}{2}}}\quad .
\end{equation}
To estimate the systematic uncertainties of our potential fits, we perform 
various types of fits, e.g. different fit-ranges in r or different  
fit-forms (3 \& 4 params. fits). In our future analysis, of the critical
temperature we will use the Sommer scale to convert our results form lattice
units to physical units. 

\section{The Critical Temperature}
\begin{figure}
\begin{minipage}{.48\textwidth}
\begin{center}
\includegraphics[width=.99\textwidth]{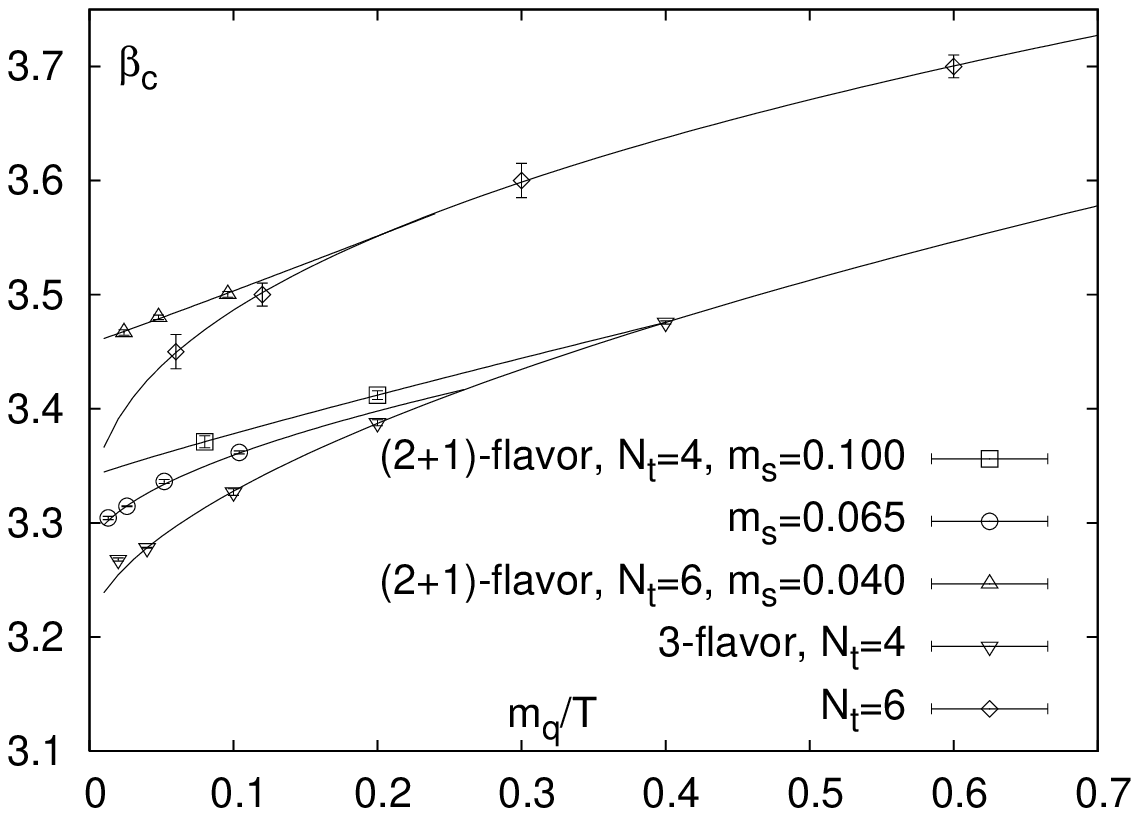}
\end{center}
\end{minipage}
\begin{minipage}{.48\textwidth}
\begin{center}
\includegraphics[width=.99\textwidth]{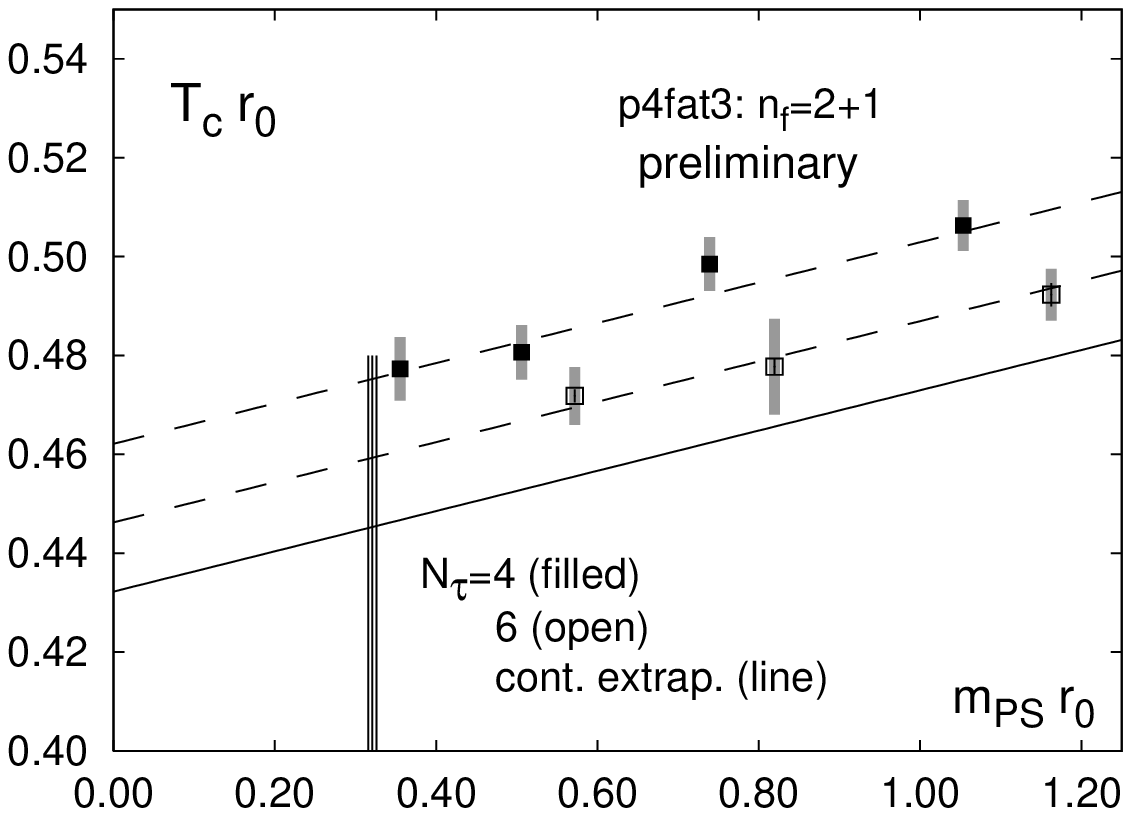}
\end{center}
\end{minipage}
\caption{The critical coupling as a function of the light quark mass
  (left) and the critical temperature as a function of the pion mass
  (right). The critical temperature as well as the pion mass are given
  in units of the Sommer scale.}\label{fig:tc}
\end{figure}

In Fig.~\ref{fig:tc} (left) we plot the critical coupling  as function
of the light quark mass. We find that $\Delta
\beta\equiv\beta_c(N_t=6)-\beta_c(N_t=4)=0.13-0.14$ is almost quark 
mass independent. This independence if the cut-off effect from quark masses
holds in leading order also for the critical Temperature as can be seen in 
Fig.~\ref{fig:tc} (right). 
Here we show $T_c$ in units of the Sommer scale
as function of the pion mass $m_{PS}$ (also in units of the Sommer scale).
We perform a combined chiral and continuum extrapolation of $T_c$ by
using the ansatz
\begin{equation}
T_c r_0= \left[T_cr_0\right]^{\rm chiral}_{\rm cont}+b_1(m_\pi r_0)+b_2/N_t^2
\quad ,
\end{equation}
where $b_1$ and $b_2$ are free fit parameters. This extrapolation leads to
critical temperature in the chiral and continuum limit of $T_c\approx
185$~MeV. The linear extrapolation in quark mass is on the one hand suggested
by the data, on the other hand it is validated by the fact that one expects a
critical point in the chiral limit which is in the $O(4)$-universality class. 
The leading order in the scaling behavior of the transition temperature is then
given by $T_c\sim (m_{PS})^{1.1}$, which is sufficiently close to a
linear scaling behavior. 

\section*{Acknowledgments}
I thank all members of the RBC-Bielefeld Collaboration for helpful
discussions and for giving me the opportunity to present our new
preliminary results prior to publication. This work has been 
supported by the the U.S. Department of Energy under contract DE-AC02-98CH1-886.


\section*{References}
\begin{thebibliography}{99}
\bibitem{Heller:1999xz}
  U.~M.~Heller, F.~Karsch and B.~Sturm,
  \Journal{\PRD}{60}{114502}{1999}.
\bibitem{Karsch:2000ps}
  F.~Karsch, E.~Laermann and A.~Peikert,
  \Journal{\PLB}{478}{447}{2000}.
\bibitem{Karsch:2000kv}
  F.~Karsch, E.~Laermann and A.~Peikert,
  \Journal{\NPB}{605}{579}{2001}.
\bibitem{p4fat7_1}
  C.~Jung [RBC-Bielefeld Collaboration],
  \Journal{\POS}{LAT2005}{150}{2005}.
\bibitem{p4fat7_2}
  M.~Cheng [RBC-Bielefeld Collaboration], 
  \Journal{\POS}{LAT2005}{045}{2005}.
\bibitem{Ferrenberg:1988yz}
  A.~M.~Ferrenberg and R.~H.~Swendsen,
  \Journal{\PRL}{61}{2635}{1988}.
\end{thebibliography}
\end{document}